\documentclass[runningheads]{llncs}
\usepackage[T1]{fontenc}
\usepackage{graphicx}
\usepackage{subfig}
\usepackage{amsmath}
\usepackage{amssymb}
\usepackage{tikz}
\usepackage{multirow}
\usepackage{standalone}
\usepackage{array}
\usepackage[pagebackref=true,breaklinks=true,letterpaper=true,colorlinks,bookmarks=false]{hyperref}
\usepackage{authblk} % Support for footnote style author/affiliation

\begin{document}
\bibliographystyle{splncs04}

\title{Region-guided CycleGANs for Stain Transfer in Whole Slide Images}

\author{Joseph Boyd\inst{1}\and
Irène Villa\inst{2} \and
Marie-Christine Mathieu\inst{2} \and
Eric Deutsch\inst{2}\and
Nikos Paragios\inst{3} \and
Maria Vakalopoulou\thanks{These authors contributed equally to this work.}\inst{1}\and
Stergios Christodoulidis$^*$\inst{1}}
%index{Boyd, Joseph}
%index{Villar, Irène}
%index{Mathieu, Marie-Christine}
%index{Deutsch, Eric}
%index{Paragios, Nikos}
%index{Vakalopoulou, Maria}
%index{Christodoulidis, Stergios}

\authorrunning{J. Boyd et al.}

\institute{MICS Laboratory, CentraleSup\'elec, Universit\'e Paris-Saclay,
91190 Gif-sur-Yvette, France
\email{firstname.lastname@centralesupelec.fr} \and
Gustave Roussy Cancer Campus, 94800 Villejuif, France
\email{firstname.lastname@gustaveroussy.fr} \and
Therapanacea, 75014 Paris, France\\
\email{n.paragios@therapanacea.eu}}
\maketitle              % typeset the header of the contribution
\begin{abstract}

In whole slide imaging, commonly used staining techniques based on hematoxylin and eosin (H\&E) and immunohistochemistry (IHC) stains accentuate different aspects of the tissue landscape. In the case of detecting metastases, IHC provides a distinct readout that is readily interpretable by pathologists. IHC, however, is a more expensive approach and not available at all medical centers. Virtually generating IHC images from H\&E using deep neural networks thus becomes an attractive alternative. Deep generative models such as CycleGANs learn a semantically-consistent mapping between two image domains, while emulating the textural properties of each domain. They are therefore a suitable choice for stain transfer applications. However, they remain fully unsupervised, and possess no mechanism for enforcing biological consistency in stain transfer. In this paper, we propose an extension to CycleGANs in the form of a region of interest discriminator. This allows the CycleGAN to learn from unpaired datasets where, in addition, there is a partial annotation of objects for which one wishes to enforce consistency. We present a use case on whole slide images, where an IHC stain provides an experimentally generated signal for metastatic cells. We demonstrate the superiority of our approach over prior art in stain transfer on histopathology tiles over two datasets. Our code and model are available at \href{https://github.com/jcboyd/miccai2022-roigan}{https://github.com/jcboyd/miccai2022-roigan}.

\end{abstract}
\section{Introduction}

\begin{figure*}[t!]
\centering
\subfloat[HES]{\includegraphics[width=0.45\textwidth]{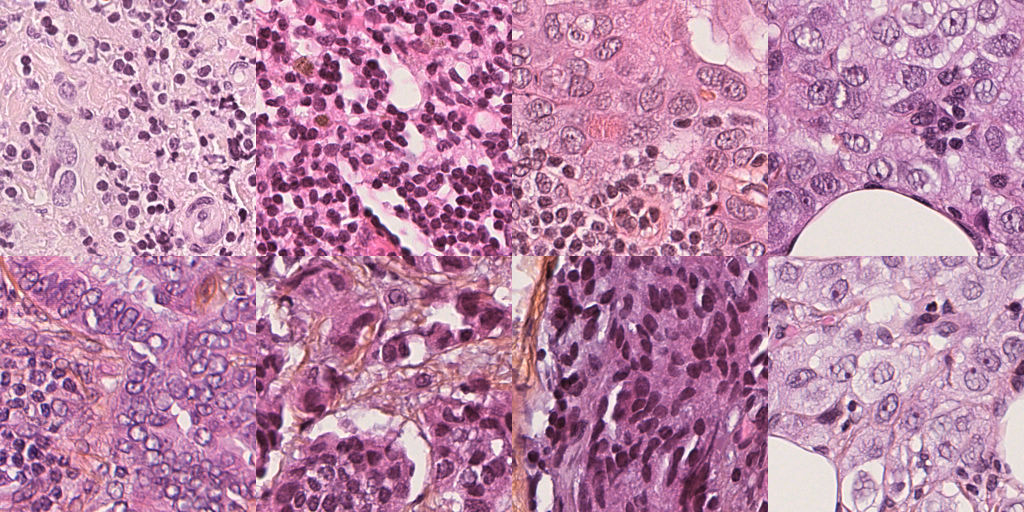}\label{subfig:hessamples}} \qquad
\subfloat[IHC]{\includegraphics[width=0.45\textwidth]{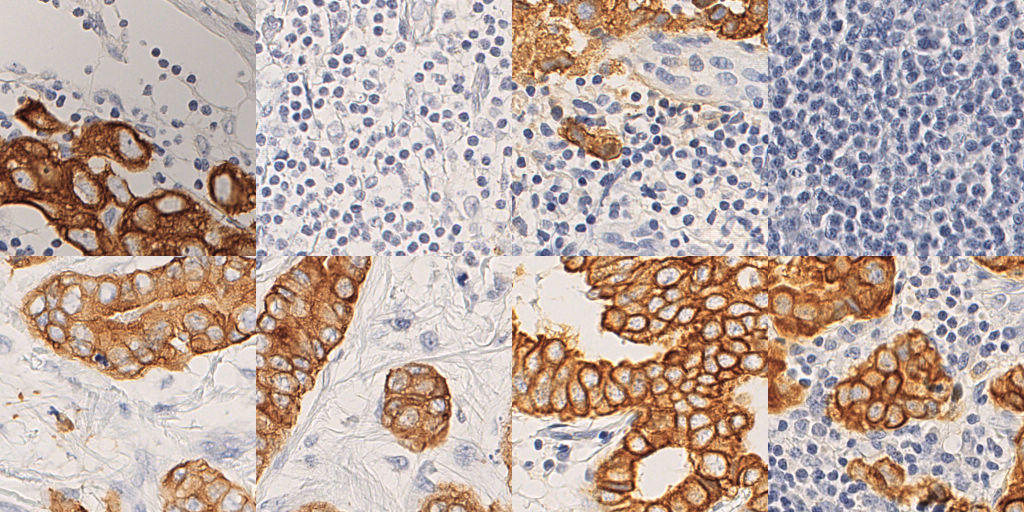}\label{subfig:ihcsamples}}
\caption{Stain transfer translates between unpaired HES (a) and IHC (b) tiles (Gustave Roussy dataset). Cancer cells visible in IHC by the golden DAB stain. }
\label{fig:datasets}
\end{figure*}

The use of histopathological whole slide images (WSI) is considered the gold standard for the diagnosis and prognosis of cancer patients. These slides contain biopsies of pathological tissue from patients and are typically stained in order to highlight different tissue structures. One of the most common tissue staining protocols is hematoxylin and eosin staining (H\&E), sometimes augmented with saffron (HES) (Figure \ref{subfig:hessamples}). Under the H\&E protocol, the hematoxylin component stains cell nuclei with a purple-blue color, while the eosin component stains the extracellular matrix and cytoplasm in pink~\cite{suvarna2018bancroft}. Another common staining is based on immunohistochemistry (IHC) (Figure \ref{subfig:ihcsamples}). This staining involves the process of selectively identifying proteins in cells and highlighting them using a chromogen. One of the most common IHC stains is based on hematoxylin together with the diaminobenzidine (DAB) chromogen forming an H-DAB stain. Under this protocol, different staining configurations can be used, targeting different cell proteins (e.g., Ki67, HER2, hormone receptors). The resulting stained tissue can provide a distinct readout, for example with AE1/AE3, in which DAB localises on cancer cell membranes. In the case of breast cancer, the presence of metastatic cells in the axillary lymph nodes can greatly affect the prognosis of a patient. Lymph node status, assessed through the WSI inspection of dissected sentinel or axillary lymph nodes, is therefore a crucial diagnostic readout. In cases where the diagnosis is difficult to make (e.g. micro-metastasis or isolated tumor cells) IHC-stained slides that highlight cancer cells are necessary.

With the recent advances in deep learning, there is a growing interest in automatically processing digitised slides. We present a methodology to computationally transfer from H\&E to IHC WSIs. Our main contribution is a region-based discriminator network within a CycleGAN framework, which utilises automatically extracted regions of interest to improve stain localisation. Unlike fully supervised schemes, our method learns from unpaired H\&E and IHC slides and produces a stain transfer that can serve as a soft segmentation for metastasis detection. Experiments on two datasets illustrate the success of our method, while demonstrating that baseline models are unable to reliably localise the DAB stain and, consequently, cancer cells. To the best of our knowledge, this is the first work to propose such a region-guided CycleGAN, as well as the first to provide robust models for stain transfer on the WSI level. Synthesised IHC slides show the potential of our method as a clinical visualisation tool, as well as in metastasis segmentation pipelines for diagnosis.

% * Describe stain transfer/how it differs from stain normalisation (even though similar models e.g. CycleGANs are used for both).

% * Motivations for HES -> IHC transfer

% * SOME DESCRIPTION ON THE NATURE OF HES and IHC Diaminobenzidine (DAB)


\section{Related Work}

Generative adversarial networks (GANs) are a popular family of generative models in histopathology image analysis~\cite{tschuchnig2020generative}. Various types of GANs have been applied to histopathology images for stain transfer~\cite{xu2019gan}, stain normalisation~\cite{rana2018computational,bentaieb2017adversarial,de2018stain}, cell segmentation~\cite{hou2019robust}, data augmentation~\cite{claudio2021pathologygan} and representation learning~\cite{boyd2021self}. A widely used variant of GANs are image-to-image translation networks, such as \texttt{pix2pix}~\cite{zhu2017unpaired}, which learns to translate between image \emph{pairs}, and CycleGANs~\cite{isola2017image}, which relax the pairing assumption to enable \emph{unpaired} and fully unsupervised image-to-image translation. CycleGANs have been used for stain transfer before, notably in \cite{xu2019gan}, which augments the CycleGAN training criterion with additional priors, so as to guide an otherwise unsupervised model.

Region of interest (RoI) information is one possible enrichment to a generative learning task, in conjunction with attention mechanisms. In \cite{savioli2018generative}, a CoGAN-inspired setup~\cite{liu2016coupled} is used for joint generation of global and RoI images. In~\cite{ouyang2018pedestrian}, pedestrians are edited into predefined scenes using \texttt{pix2pix}, with spatial pyramid pooling in the discriminator for direct scrutiny. Pre-trained R-CNN object detection systems have been used to propose regions during GAN training in~\cite{huang2019realistic}, or as a feature extractor in object-driven GANs~\cite{li2019object}. In contrast, we modify the GAN discriminator itself, and leverage automatically derived RoI data.

\section{Method}
\label{sec:methods}
%\subsection{CycleGANs for stain transfer}
CycleGANs~\cite{zhu2017unpaired} lend themselves to the task of stain transfer between unpaired histopathology tiles. These are unsupervised models incorporating two GAN generators, $G_{XY} : X \to Y$ and $G_{YX} : Y \to X$ for \emph{unpaired} image domains $X$ and $Y$. In an application of stain transfer between two stains, each stain represents a different domain. Although both generators are trained, in our case only one direction of transfer is desired (HES $\to$ IHC), and the other generator may be discarded after training. CycleGAN training is performed according to,

\begin{align}
\min_{G_{XY}, G_{YX}}\max_{D_X, D_Y} &{}\mathcal{L}_{CG} = \mathcal{L}_{GAN}(D_Y, G, X, Y) + \mathcal{L}_{GAN}(G_{XY}, D_Y, X, Y)\label{eq:cycle_gan}\\
&+ \lambda_{CYC}\mathcal{L}_{CYC}(G_{XY}, G_{YX}, X, Y) \notag + \lambda_{ID}\mathcal{L}_{ID}(G_{XY}, G_{YX}, X, Y), \notag
\end{align}

which combines least squares adversarial losses~\cite{mao2017least} $\mathcal{L}_{GAN}$, with PatchGAN discriminators $D_X$ and $D_Y$, along with a cycle-consistency loss $\mathcal{L}_{CYC}$ to maintain pixel-wise consistency back and forth between domains, identity function losses $\mathcal{L}_{ID}$ for stability, and $\lambda_{CYC}$ and $\lambda_{ID}$ are hand-tuned weights.

% Pixel-wise consistency is enforced with a cycle consistency loss,

% \begin{equation}
% \mathcal{L}_{CYC}(G_{XY}, G_{YX}, X, Y) = \mathcal{L}_{1}(G_{YX}(G_{XY}(X), X)) + \mathcal{L}_{1}(G_{XY}(G_{YX}(Y), Y))
% \label{eq:cycle_consistency}
% \end{equation}

% that is, the mapping of an image to the other domain and back. Such a process must approximately reconstruct the original image. For reasons of stability, this loss is combined with identity function losses,

% \begin{equation}
% \mathcal{L}_{ID}(G_{XY}, G_{YX}, X, Y) = \mathcal{L}_{1}(G_{XY}(Y), Y) + \mathcal{L}_{1}(G_{YX}(X), X)
% \label{eq:identity_loss}
% \end{equation}

% Equations \ref{eq:cycle_consistency} and \ref{eq:identity_loss} are combined with adversarial losses into the final CycleGAN loss function and trained as,

% where the $\mathcal{L}_{GAN}$ are typically least-square adversarial losses~\cite{mao2017least}, rather than the more standard Jensen-Shannon adversarial loss~\cite{goodfellow2014generative}, and $\lambda_{CYC}$, $\lambda_{ID}$ are hand-tuned weights.

\subsection{Region of interest discrimination}
\label{subsec:roi_disc}

\begin{figure*}[t]%
    \centering
    %\subfloat[Region-based discriminator]{\includestandalone[height=.15\linewidth]{tikz/roi-gan}}
    \subfloat[Patch-based discriminator]{
    \tikzset{every picture/.style={line width=0.25pt}} %set default line width to 0.75pt   
\begin{tikzpicture}[x=0.75pt,y=0.75pt,yscale=-1,xscale=1]

%Straight Lines [id:da7962990769140612] 
\draw    (100,4) -- (100,76) ;
%Straight Lines [id:da2716994179083623] 
\draw    (28,76) -- (100,76) ;
%Straight Lines [id:da583088380758828] 
\draw    (60,4) -- (60,76) ;
%Shape: Square [id:dp9180809174484449] 
\draw  [color={rgb, 255:red, 255; green, 0; blue, 0 }  ,draw opacity=1 ][line width=0.75]  (200,28) -- (208,28) -- (208,36) -- (200,36) -- cycle ;
%Shape: Square [id:dp37442755183112564] 
\draw  [color={rgb, 255:red, 255; green, 0; blue, 0 }  ,draw opacity=1 ][line width=0.75]  (208,36) -- (216,36) -- (216,44) -- (208,44) -- cycle ;
%Straight Lines [id:da5891494102822498] 
\draw    (200,52) -- (212,52) ;
%Straight Lines [id:da5113013478186778] 
\draw    (212,52) -- (224,52) ;
%Straight Lines [id:da8831105229052274] 
\draw    (224,40) -- (224,52) ;
%Straight Lines [id:da8654004481948433] 
\draw    (224,28) -- (224,40) ;
%Straight Lines [id:da22183733068395883] 
\draw    (68,4) -- (100,4) ;
%Straight Lines [id:da11266851998230243] 
\draw    (28,44) -- (28,76) ;
%Straight Lines [id:da4064331764819059] 
\draw    (28,36) -- (100,36) ;
%Straight Lines [id:da533644070645898] 
\draw    (68,44) -- (100,44) ;
%Straight Lines [id:da09736143883363535] 
\draw    (28,52) -- (100,52) ;
%Straight Lines [id:da6675262148315964] 
\draw    (28,60) -- (44,60) ;
%Straight Lines [id:da3958699437187224] 
\draw    (32,4) -- (32,76) ;
%Straight Lines [id:da562717744413308] 
\draw    (36,4) -- (36,76) ;
%Straight Lines [id:da1987615330908774] 
\draw    (40,4) -- (40,76) ;
%Straight Lines [id:da9723755941666333] 
\draw    (44,60) -- (44,76) ;
%Straight Lines [id:da11347053253581507] 
\draw    (48,4) -- (48,76) ;
%Straight Lines [id:da833541507362869] 
\draw    (52,4) -- (52,76) ;
%Straight Lines [id:da544062795922843] 
\draw    (56,4) -- (56,76) ;
%Straight Lines [id:da7767372459118718] 
\draw    (64,4) -- (64,76) ;
%Straight Lines [id:da04528166861689609] 
\draw    (68,44) -- (68,76) ;
%Straight Lines [id:da03616118571091287] 
\draw    (72,4) -- (72,76) ;
%Straight Lines [id:da6796805058406482] 
\draw    (76,4) -- (76,76) ;
%Straight Lines [id:da636756498704046] 
\draw    (80,4) -- (80,76) ;
%Straight Lines [id:da986507333296601] 
\draw    (84,60) -- (84,76) ;
%Straight Lines [id:da22342875006734975] 
\draw    (88,4) -- (88,76) ;
%Straight Lines [id:da49852141340892964] 
\draw    (92,4) -- (92,76) ;
%Straight Lines [id:da07691958942809685] 
\draw    (96,4) -- (96,76) ;
%Straight Lines [id:da5533689930951443] 
\draw    (28,8) -- (100,8) ;
%Straight Lines [id:da48570711631817776] 
\draw    (28,12) -- (100,12) ;
%Straight Lines [id:da9436628873971031] 
\draw    (28,16) -- (100,16) ;
%Straight Lines [id:da6339043386925132] 
\draw    (28,20) -- (44,20) ;
%Straight Lines [id:da3302627415189895] 
\draw    (28,24) -- (100,24) ;
%Straight Lines [id:da5030599608484946] 
\draw    (28,32) -- (100,32) ;
%Straight Lines [id:da41181103134549124] 
\draw    (28,68) -- (100,68) ;
%Straight Lines [id:da6891587898398969] 
\draw    (28,40) -- (100,40) ;
%Straight Lines [id:da1504397269785639] 
\draw    (28,48) -- (100,48) ;
%Straight Lines [id:da5201305841950166] 
\draw    (28,56) -- (100,56) ;
%Straight Lines [id:da5105969389943457] 
\draw    (28,64) -- (100,64) ;
%Straight Lines [id:da4494823248097465] 
\draw    (28,72) -- (100,72) ;
%Shape: Square [id:dp8285067079399483] 
\draw  [color={rgb, 255:red, 255; green, 0; blue, 0 }  ,draw opacity=1 ][line width=0.75]  (28,4) -- (68,4) -- (68,44) -- (28,44) -- cycle ;
%Straight Lines [id:da8826196537166912] 
\draw    (216,28) -- (216,36) ;
%Straight Lines [id:da3133032663582713] 
\draw    (200,36) -- (200,52) ;
%Straight Lines [id:da29476880491648105] 
\draw    (224,28) -- (208,28) ;
%Straight Lines [id:da5668323113856902] 
\draw    (224,36) -- (216,36) ;
%Straight Lines [id:da2559708309350718] 
\draw    (208,44) -- (208,52) ;
%Straight Lines [id:da6379457189599795] 
\draw    (208,44) -- (200,44) ;
%Shape: Square [id:dp6924699203245184] 
\draw  [color={rgb, 255:red, 255; green, 0; blue, 0 }  ,draw opacity=1 ][line width=0.75]  (44,20) -- (84,20) -- (84,60) -- (44,60) -- cycle ;
%Straight Lines [id:da49742622650037027] 
\draw    (44,4) -- (44,20) ;
%Straight Lines [id:da6660771300050975] 
\draw    (84,20) -- (100,20) ;
%Straight Lines [id:da2970695578752405] 
\draw    (28,28) -- (100,28) ;
%Straight Lines [id:da8990463033264174] 
\draw    (84,60) -- (100,60) ;
%Straight Lines [id:da411869241958737] 
\draw    (84,4) -- (84,20) ;
%Straight Lines [id:da5831459334818176] 
\draw    (224,44) -- (216,44) ;
%Straight Lines [id:da8324039524070977] 
\draw    (216,44) -- (216,52) ;
%Straight Lines [id:da9796671935325555] 
\draw    (108,76) -- (192,52.72) ;
%Straight Lines [id:da2928369389529264] 
\draw    (108,76) -- (108,4) ;
%Straight Lines [id:da37839787058739993] 
\draw    (108,4) -- (192,28) ;
%Straight Lines [id:da307893555390099] 
\draw    (192,52.72) -- (192,28) ;

% Text Node
\draw (226,40) node [anchor=west] [inner sep=0.75pt]  [font=\scriptsize]  {$h$};
% Text Node
\draw (212,55.4) node [anchor=north] [inner sep=0.75pt]  [font=\scriptsize]  {$w$};
\end{tikzpicture}
\label{fig:patchvsroi_patch}
    }%
    \qquad
    \subfloat[Region-based discriminator]{ %set default line width to 0.75pt
\tikzset{every picture/.style={line width=0.25pt}}
\begin{tikzpicture}[x=0.75pt,y=0.75pt,yscale=-1,xscale=1]

%Shape: Square [id:dp2262004570440146] 
\draw  [color={rgb, 255:red, 255; green, 0; blue, 0 }  ,draw opacity=1 ][line width=0.75]  (182,42) -- (190,42) -- (190,50) -- (182,50) -- cycle ;
%Shape: Square [id:dp43304025375039634] 
\draw  [color={rgb, 255:red, 255; green, 0; blue, 0 }  ,draw opacity=1 ][line width=0.75]  (182,58) -- (190,58) -- (190,66) -- (182,66) -- cycle ;
%Shape: Square [id:dp24830085416443415] 
\draw  [color={rgb, 255:red, 255; green, 0; blue, 0 }  ,draw opacity=1 ][line width=0.75]  (182,50) -- (190,50) -- (190,58) -- (182,58) -- cycle ;
%Shape: Square [id:dp8755927469886827] 
\draw  [color={rgb, 255:red, 255; green, 0; blue, 0 }  ,draw opacity=1 ][line width=0.75]  (182,34) -- (190,34) -- (190,42) -- (182,42) -- cycle ;
%Shape: Square [id:dp09180814397537407] 
\draw  [color={rgb, 255:red, 255; green, 0; blue, 0 }  ,draw opacity=1 ][line width=0.75]  (182,26) -- (190,26) -- (190,34) -- (182,34) -- cycle ;

%Straight Lines [id:da3903930196241874] 
\draw    (90,82) -- (142,66) ;
%Straight Lines [id:da8311248501409504] 
\draw    (90,82) -- (90,10) ;
%Straight Lines [id:da44706422457331096] 
\draw    (90,10) -- (142,26) ;
%Straight Lines [id:da1784851919894208] 
\draw    (142,66) -- (142,26) ;
%Straight Lines [id:da397413172774565] 
\draw    (174,66) -- (174,45.33) -- (174,26) ;
%Straight Lines [id:da6540936655891519] 
\draw    (158,66) -- (158,26) ;
%Straight Lines [id:da07842600577648362] 
\draw    (158,26) -- (142,26) ;
%Straight Lines [id:da27477993260354916] 
\draw    (174,26) -- (158,26) ;
%Straight Lines [id:da20987536923713312] 
\draw    (158,66) -- (142,66) ;
%Straight Lines [id:da21512791472567372] 
\draw    (174,66) -- (158,66) ;
%Straight Lines [id:da74863775934807] 
\draw    (82,10) -- (82,82) ;
%Straight Lines [id:da7201354283731033] 
\draw    (10,82) -- (82,82) ;
%Straight Lines [id:da4010766703286155] 
\draw    (42,10) -- (42,82) ;
%Straight Lines [id:da8856326730415931] 
\draw    (10,10) -- (10,82) ;
%Straight Lines [id:da5244582274457423] 
\draw    (14,10) -- (14,82) ;
%Straight Lines [id:da28908566859884854] 
\draw    (18,10) -- (18,82) ;
%Straight Lines [id:da2925949087239218] 
\draw    (26,10) -- (26,82) ;
%Straight Lines [id:da7270378409831928] 
\draw    (22,10) -- (22,82) ;
%Straight Lines [id:da3997415571361401] 
\draw    (34,10) -- (34,82) ;
%Straight Lines [id:da26281799825194085] 
\draw    (38,10) -- (38,82) ;
%Straight Lines [id:da1268351643770943] 
\draw    (46,10) -- (46,82) ;
%Straight Lines [id:da9432229116386972] 
\draw    (54,10) -- (54,82) ;
%Straight Lines [id:da30439736629336234] 
\draw    (58,10) -- (58,82) ;
%Straight Lines [id:da5859282917281765] 
\draw    (62,10) -- (62,82) ;
%Straight Lines [id:da34587344904903106] 
\draw    (66,10) -- (66,82) ;
%Straight Lines [id:da4219209068393145] 
\draw    (70,10) -- (70,82) ;
%Straight Lines [id:da3580337487256966] 
\draw    (74,10) -- (74,82) ;
%Straight Lines [id:da03733178645378943] 
\draw    (78,10) -- (78,82) ;
%Straight Lines [id:da7162374406546654] 
\draw    (10,78) -- (82,78) ;
%Straight Lines [id:da4014120748385289] 
\draw    (10,74) -- (82,74) ;
%Straight Lines [id:da5684175781605295] 
\draw    (10,70) -- (82,70) ;
%Straight Lines [id:da24897625778033983] 
\draw    (10,66) -- (82,66) ;
%Straight Lines [id:da9362870190523165] 
\draw    (10,62) -- (82,62) ;
%Straight Lines [id:da7991905255404703] 
\draw    (10,58) -- (82,58) ;
%Straight Lines [id:da6567669064132486] 
\draw    (10,54) -- (82,54) ;
%Straight Lines [id:da8975986711683044] 
\draw    (10,50) -- (82,50) ;
%Straight Lines [id:da36348558328116676] 
\draw    (10,46) -- (82,46) ;
%Straight Lines [id:da05519916895765287] 
\draw    (10,42) -- (82,42) ;
%Straight Lines [id:da4991207657966854] 
\draw    (10,38) -- (82,38) ;
%Straight Lines [id:da2637527369969177] 
\draw    (10,34) -- (82,34) ;
%Straight Lines [id:da2829668226635931] 
\draw    (10,30) -- (82,30) ;
%Straight Lines [id:da6153177400783575] 
\draw    (10,26) -- (82,26) ;
%Straight Lines [id:da32619432641163537] 
\draw    (10,22) -- (82,22) ;
%Straight Lines [id:da347160238053722] 
\draw    (10,18) -- (82,18) ;
%Straight Lines [id:da6029779983974135] 
\draw    (10,14) -- (82,14) ;
%Straight Lines [id:da10676922914425468] 
\draw    (10,10) -- (82,10) ;
%Straight Lines [id:da6869658799606565] 
\draw    (30,10) -- (30,82) ;
%Straight Lines [id:da24578243127594446] 
\draw    (50,10) -- (50,82) ;

%Flowchart: Process [id:dp09508504614715962] 
\draw  [color={rgb, 255:red, 255; green, 0; blue, 0 }  ,draw opacity=1 ][fill={rgb, 255:red, 255; green, 255; blue, 255 }  ,fill opacity=0 ][line width=0.75]  (46,42) -- (74,42) -- (74,74) -- (46,74) -- cycle ;
%Flowchart: Process [id:dp317364739872674] 
\draw  [color={rgb, 255:red, 255; green, 0; blue, 0 }  ,draw opacity=1 ][fill={rgb, 255:red, 255; green, 255; blue, 255 }  ,fill opacity=0 ][line width=0.75]  (66,18) -- (74,18) -- (74,34) -- (66,34) -- cycle ;
%Flowchart: Process [id:dp9008437509822393] 
\draw  [color={rgb, 255:red, 255; green, 0; blue, 0 }  ,draw opacity=1 ][fill={rgb, 255:red, 255; green, 255; blue, 255 }  ,fill opacity=0 ][line width=0.75]  (18,42) -- (38,42) -- (38,54) -- (18,54) -- cycle ;
%Flowchart: Process [id:dp2388779120278659] 
\draw  [color={rgb, 255:red, 255; green, 0; blue, 0 }  ,draw opacity=1 ][fill={rgb, 255:red, 255; green, 255; blue, 255 }  ,fill opacity=0 ][line width=0.75]  (18,62) -- (38,62) -- (38,74) -- (18,74) -- cycle ;
%Flowchart: Process [id:dp24667136491274644] 
\draw  [color={rgb, 255:red, 255; green, 0; blue, 0 }  ,draw opacity=1 ][fill={rgb, 255:red, 255; green, 255; blue, 255 }  ,fill opacity=0 ][line width=0.75]  (18,18) -- (58,18) -- (58,34) -- (18,34) -- cycle ;

% Text Node
\draw (186,69.4) node [anchor=north] [inner sep=0.75pt]  [font=\scriptsize]  {$k$};
% Text Node
\draw (116,46) node  [font=\scriptsize]  {$f$};
% Text Node
\draw (150,46) node  [font=\scriptsize]  {$\rho $};
% Text Node
\draw (166,46) node  [font=\scriptsize]  {$d$};

\end{tikzpicture}\label{fig:patchvsroi_roi}}
\caption{Conventional PatchGAN discriminates over a regular grid of overlapping receptive fields (a) while a region-based discriminator may discriminate an image on arbitrarily-sized regions (b).}\label{Fig1}
\label{fig:patchvsroi}
\end{figure*}
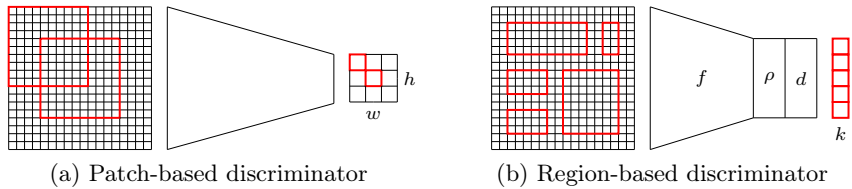

Image photo-realism is enforced in a CycleGAN by a ``PatchGAN'' discriminator,

\begin{equation}
D_{patch} : X \to \{0, 1\}^{h\times w},
\end{equation}

which consists of a sequence of strided convolutions producing a $h\times w$ grid of outputs, with each grid element discriminating an overlapping patch of the inputs, as in Figure \ref{fig:patchvsroi_patch}. The output grid is compared element-wise with a grid of ground truth labels. As will be quantified in Section \ref{sec:results}, this purely unsupervised, standard CycleGAN setup fails to correctly localise DAB in a stain transfer application. This motivates a new type of discriminator, based on region of interest discrimination via a region of interest alignment (RoIAlign) layer. The RoIAlign layer was originally proposed in the Mask R-CNN object detection system~\cite{he2017mask}, and is a generalisation of the classical MaxPool layer, and a way of quantising regions of activation maps of arbitrary size into a standardised dimension. Our proposed RoI discriminator consists of feature extraction layers $f$ followed by a RoIAlign layer $\rho$, and a final discrimination layer $d$, as shown in Figure \ref{fig:patchvsroi_roi}. We formalise it as,

\begin{equation}
D_{roi} : X \times B \to \{0, 1\}^k,
\end{equation}

for image domain $X$ and bounding box domain $B$, where $k$ is the number of bounding boxes. The adversarial loss thus becomes,

\begin{equation}
\mathcal{L}_{ROI}(G, D_{roi}, X, Y) = \frac{1}{2}\mathbb{E}_{\mathbf{x}\in X}[(D_{roi}(\mathbf{x}, \mathcal{B}(\mathbf{x})) - 1)^2] + \frac{1}{2}\mathbb{E}_{\mathbf{y}\in Y}[D_{roi}(G(\mathbf{y}), \mathcal{B}(\mathbf{y}))^2]
\label{eq:roi_adversarial}
\end{equation}

that is, a least squares adversarial loss, where the operator $\mathcal{B}$ returns the set of bounding boxes of an image, provided by a previously generated object library. A pair of RoI discriminators (one for each domain) can then be trained alongside--or instead of--the PatchGAN discriminators (see Appendix A).

\subsection{Library generation for region-based discrimination}
\label{subsec:library}

To train our proposed region-based discriminator, we build a bounding box library by localising cancer and normal cells using an automatic image processing pipeline (see Appendix B). All input tiles (see Section \ref{sec:datasets}) are first decomposed into their hematoxylin-eosin-DAB components. The expert annotations available for H\&E slides identify cancerous regions that may nevertheless contain many interspersed healthy cells. Blob detection applied directly on the HES images is thus prone to false positives, precisely when precision is more important to our downstream algorithm than recall. To mitigate this effect, we apply a Laplacian of Gaussian (LoG) filter ($\sigma = \{1, 2\}$). We then construct a graph of healthy cells where connectivity is determined by a manual threshold ($25$ pixels) on Euclidean distance between detected cells. Graph connected components constitute healthy cell clusters, and the convex hull of each cluster is zeroed-out of the expert annotation. In a final step, a second pass of blob detection ($\sigma=\{10, \dots, 12\}$) is applied to detect the larger, dimmer cancer cells in the unmasked regions. In the case of IHC tiles, we instead perform an Otsu threshold on the DAB channel to isolate the cancerous sub-regions. Cancer cell nuclei feature as elliptical discontinuities in the binarised DAB stain, and can again be detected using a LoG filter ($\sigma = \{7, \dots, 10\}$). Smaller blobs ($\sigma = \{1, \dots, 3\}$) outside the segmented region are taken to be normal cells. The library, consisting of over $100$k cell bounding boxes, is then used for the training of the RoI discriminator.

\subsection{Implementation and training details}

In each experiment, the two CycleGAN generators follow the architecture of a baseline approach~\cite{xu2019gan}, itself based on a network proposed for style transfer~\cite{johnson2016perceptual}. The tile inputs are of size $256\times256$ with three (RGB) channels. The generators consist first of strided convolutions to lower the resolution. A sequence of residual blocks~\cite{he2016deep} of tunable length is then applied. We benchmark $6$ and $9$ residual blocks. Following this, fractionally-strided convolutions restore the input resolution. The PatchGAN discriminators consist of a standard sequence of five strided convolutions and an output layer. This reduces the input to a $8\times8$ patch-wise prediction. This architecture, hereafter referred to as CycleGAN $8\times8$, in reference to the discriminator output size, is trained according to Equation~\ref{eq:cycle_gan}. The baselines and proposed model are all modifications of this core baseline.

Given that our proposed discriminators have a selective receptive field, we control for the receptive field size of the PatchGAN discriminators by modifying the stride of their convolutions. We modify the stride of the first layer from $2$ to $1$ in baseline CycleGAN $16\times16$, doubling the patch output size, and the first two layers in baseline CycleGAN $32\times32$, doubling again. We also implement the conditional CycleGAN model~\cite{xu2019gan}. This introduces two additional networks of its own, which are used to classify tiles in each domain. Here, Equation \ref{eq:cycle_gan} is supplemented with classification and cycle classification losses for the new networks, as well as novel photo-realism and structural similarity losses. Exceptionally, due to the speed and memory constraints of the photo-realism loss, we train this baseline with batch size $2$. In addition, we test a plain conditional CycleGAN without these additional losses.

For our proposed model, the RoI discriminator resembles the PatchGAN, with four stride-$2$ feature convolutions $f$, followed by a RoIAlign layer $\rho$, and a final discrimination layer $d$. Without loss of generality, we restrict cell bounding box dimension to $48\times48$, centered on the library nuclei. Although the cell populations in each tile are often imbalanced, we sample balanced numbers of positives and negatives with replacement, whenever available, equaling  $k = 8$ bounding boxes per tile, and we skip tiles containing no library cells. All model weights are randomly initialised and all models use the Adam optimiser~\cite{kingma2014adam} with maximum learning rate $2$-e4 and $(\beta_0, \beta_1) = (0.5, 0.999)$. Models were trained for $20$ epochs with a batch size of $8$ randomly sampled tiles, consuming roughly two hours of processing time per model on an NVIDIA Tesla V100 GPU.

\section{Datasets}
\label{sec:datasets}

% \begin{figure}%
% \centering
% \subfloat[HES]{\includegraphics[height=0.45\textwidth]{img/inputs_mosaic.png}} \qquad
% \subfloat[IHC]{\includegraphics[height=0.45\textwidth]{img/ihc_mosaic.png}}%\qquad
% % \subfloat[Region-guided CycleGAN]{\includestandalone[height=0.45\textwidth]{tikz/cyclegan}}
% \caption{Sample HES (a) and IHC (b) tiles from private dataset. Model overview (c) illustrates forward pass of $G_{XY}$, with $\mathcal{B}$ operator supplying $D_{ROI}$ with bounding box data $(x, y, $width, height$)$. Note that $G_{YX}$ follows the same structure, with its own pair of discriminators (not pictured).}\label{Fig1}
% \end{figure}

To benchmark our method, we conducted experiments on two breast cancer datasets. One dataset was provided by the Gustave Roussy (GR) Institute consisting of $205$ WSIs, corresponding to sentinel lymph node sections for patients of breast cancer. Each lymph node was imaged with both HES and cytokeratin AE1/AE3 IHC. Although the two stainings are performed on closely situated sections of tissue, the lack of precise alignment implies an unpaired image dataset. Metastatic regions of the HES slides were annotated with bounding contours by two expert pathologists. The dataset comprises cases of micro- and macro-metastases (resp. $0.2$mm-$2$mm and $\geq 2$mm tumours), as well as negative cases. Non-overlapping tiles of size $256\times256$ pixels were extracted at a magnification of $20$x from the segmented tissue regions of WSIs using CLAM~\cite{lu2021data}. For HES slides, the expert annotation is used to extract balanced samples of positive (metastatic tissue) and negative (normal tissue) tiles whereas, for IHC, the thresholded DAB stain substitutes as an experimentally-generated annotation.

CAMELYON16~\cite{litjens20181399} consists of $1399$ hematoxylin-eosin-stained (H\&E) slides. Among these, $111$ slides were annotated in a fashion similar to the GR dataset. We processed and sampled tiles from these slides in the same way, yielding a dataset of $12000$ tiles. Note that this dataset features a slightly different imaging modality, and furthermore contains no IHC slides. Nevertheless, we found that IHC tiles from the GR dataset were fully compatible with CAMELYON16 during model training (provided the imaging magnification), demonstrating that one can combine datasets of different origin for stain transfer.

\section{Experimental results}
\label{sec:results}

\subsection{Tile-level quantitative results}
\label{subsec:tilelevel}

Our primary means of evaluating model performance is with the expert H\&E annotations. This provides a good first approximation to the locations of the cancer cells, and where the DAB should appear once the stain transfer has been performed, even though annotation contours often bisect healthy tissue regions or otherwise contain normal cells. For each of our trained generators, we first produce the IHC stain for each of a set of $500$ held-out test H\&E tiles. We visualise samples of these in Figure \ref{fig:datasets} for both the proposed model and the baseline CycleGAN $8\times8$, alongside the annotation. One may observe the accuracy of our proposed model in localising the DAB stain. Although we observe a weak correlation between positive tiles and the presence of DAB, the baseline CycleGAN systematically misplaces the DAB stain. This localisation problem was observed across all competing models. We hypothesise our proposed model profits from the object-level supervision, with our proposed RoI discriminator performing discrimination directly centered on cells.

\begin{figure*}[t!]%
\centering
\subfloat[Input images]{\includegraphics[width=0.45\textwidth]{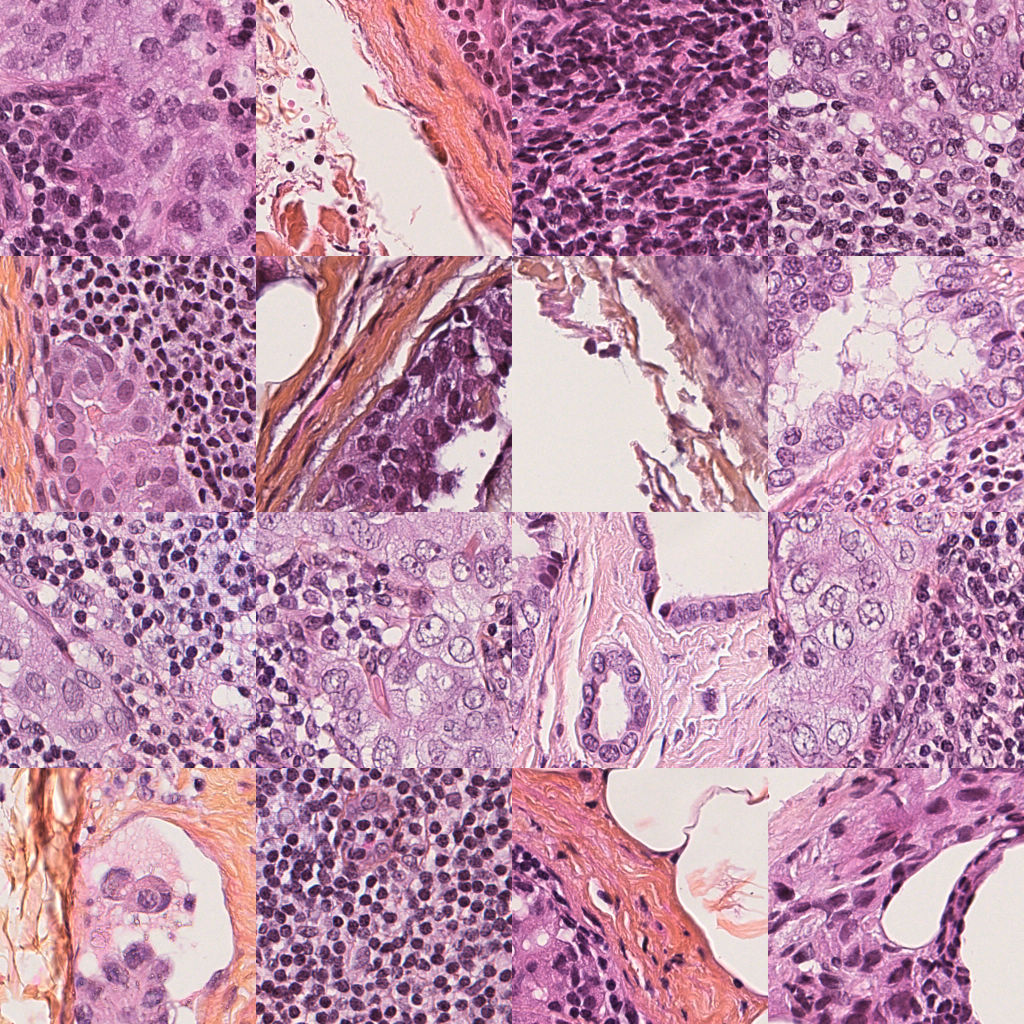}} \qquad
\subfloat[Annotation]{\includegraphics[width=0.45\textwidth]{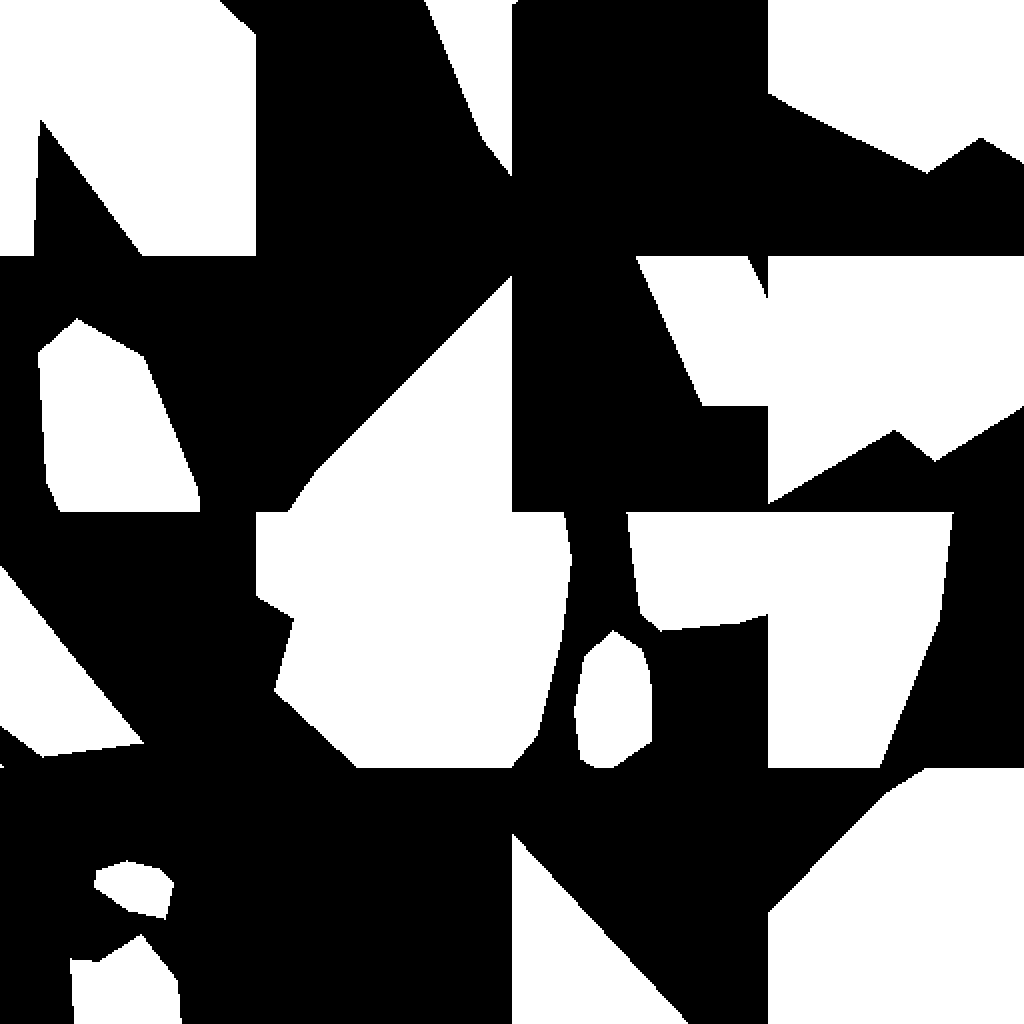}} \qquad
\subfloat[CycleGAN]{\includegraphics[width=0.45\textwidth]{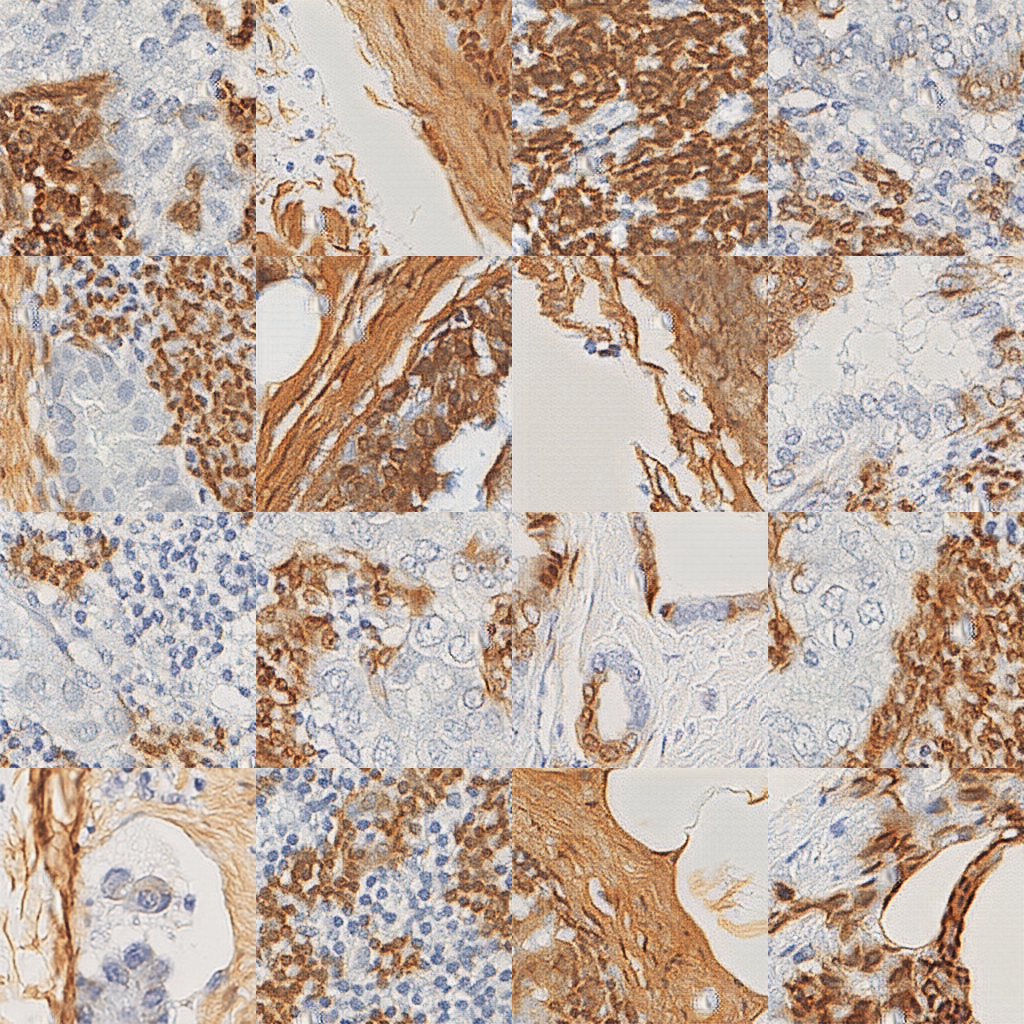}} \qquad
\subfloat[Region-guided CycleGAN]{\includegraphics[width=0.45\textwidth]{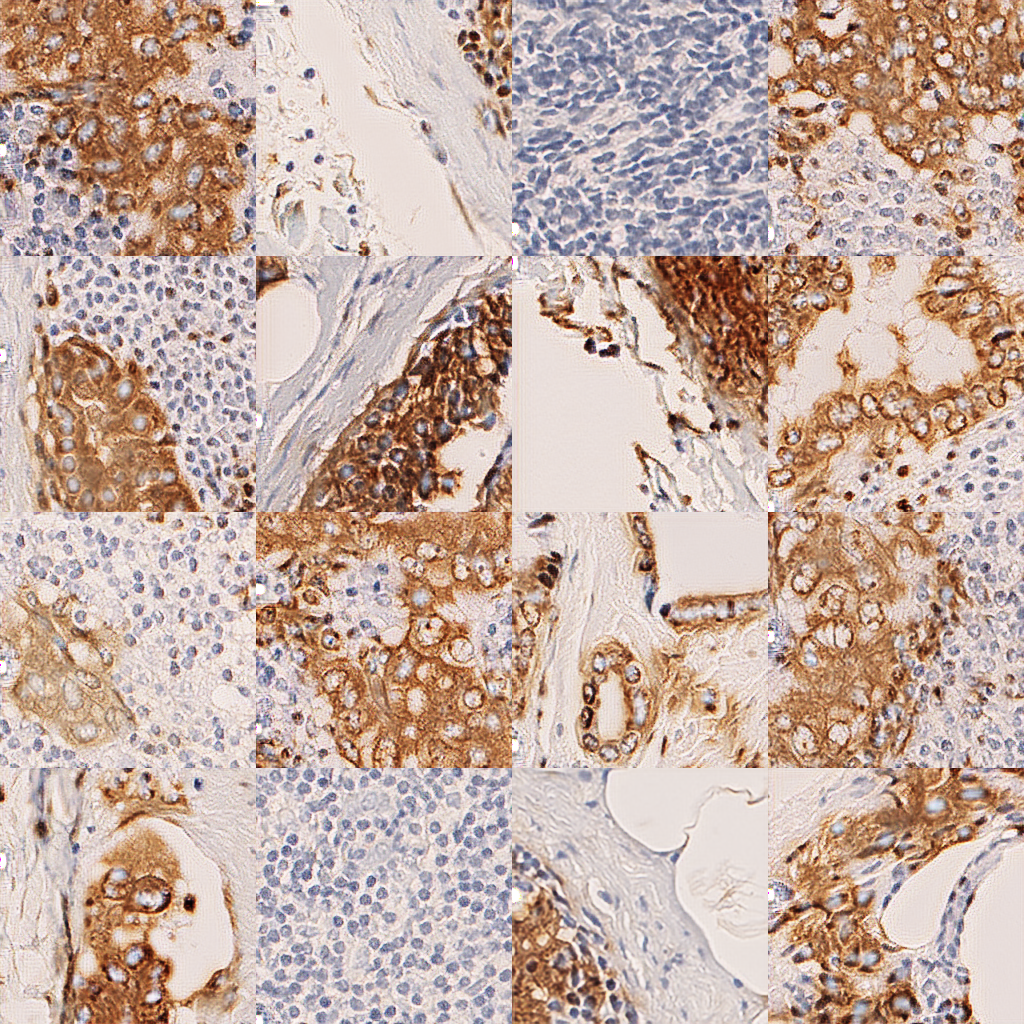}}
\caption{Sample H\&E tiles from the GR dataset (a); with corresponding ground truth annotation masks (b); baseline CycleGAN stain transfer (c); and proposed region-guided stain transfer (d).}
\label{fig:datasets}
\end{figure*}

In pursuing a quantitative evaluation, we obtain a binary mask for each synthesised DAB stain in the same manner as was performed in Section \ref{subsec:library} for real IHC tiles. This mask is then compared with the ground truth annotation by DICE similarity and balanced accuracy (BAC), and we report the mean and standard deviation for each metric across all test tiles in Table \ref{table:results}. We emphasise the ground truth is approximate, as it is intended to capture metastatic regions at a macro level, and often does not exclude healthy sub-regions. For each model, we select the best performing instance in a grid search over the number of generator residual blocks $(6, 9)$, and $\lambda_{cyc} = \{1, 10, 30\}$. Models generally performed better with $6$ residual blocks, and with the standard $\lambda_{CYC} = 10$. Our proposed model performs significantly better for $\lambda_{CYC} = 30$, however, seemingly to compensate for the weight of the additional discriminators. We nevertheless note the superiority of our proposed model over all baselines, with $0.536$ DICE and $0.634$ BAC on the GR dataset, and $0.533$ DICE and $0.615$ BAC on CAMELYON16. Among the baselines, we observe that changing the discriminator outputs ($16\times16$ and $8\times8$) can have a positive effect on BAC, but ultimately falls short on DICE. On the other hand, the conditional CycleGANs do not show improvement over the plain baselines. Finally, we observe that our high performance is sustained on CAMELYON16, even though in the absence of a IHC data, GR IHC tiles, of different clinical origin, were used as surrogate.

\begin{table}[t!]
\begin{center}
\begin{tabular}{ |c|c|c|c|c| } 
\hline
& \multicolumn{2}{c|}{GR Dataset} & \multicolumn{2}{c|}{CAMEYLON16} \\
  \hline
  Model & DICE & BAC & DICE & BAC \\ 
  \hline
  CycleGAN ($8 \times 8$) & $0.442\pm0.370$ & $0.490\pm0.103$ & $0.415\pm0.372$ & $0.501\pm0.091$ \\ 
  \hline
  CycleGAN ($16 \times 16$) & $0.365\pm0.330$ & $0.520\pm0.125$ & $0.388\pm0.349$ & $0.502\pm0.091$ \\ 
  \hline
  CycleGAN ($32 \times 32$) & $0.384\pm0.324$ & $0.511\pm0.130$ & $0.403\pm0.358$ & $0.503\pm0.100$ \\ 
  \hline
  Conditional CycleGAN & $0.341\pm0.328$ & $0.488\pm0.120$ & $0.412\pm0.366$ & $0.501\pm0.100$ \\ 
  \hline
  Xu et al. (2019)~\cite{xu2019gan} & $0.325\pm0.310$ & $0.474\pm0.107$ & $0.260\pm0.257$ & $0.498\pm0.115$ \\ 
  \hline
  Proposed & $\mathbf{0.536\pm0.360}$ & $\mathbf{0.634\pm0.221}$ & $\mathbf{0.533\pm0.371}$ & $\mathbf{0.615\pm0.197}$ \\
  \hline
\end{tabular}
\caption{DICE similarity and balanced accuracy (BAC) for all baselines and proposed model for GR dataset and CAMELYON16. Each entry reads mean $\pm$ std. Best results in bold.  \label{table:results}}
\end{center}
\end{table}

\subsection{Slide-level qualitative results}

We further explore the capabilities of our method by generating whole slide outputs, as shown for a macrometastatic case (Appendix C, Figure 3a, b, and c) and negative case (Appendix C, Figure 3d, e, and f) from the GR dataset. One may observe that the HES and IHC ground truth slides are unregistered. Here, the model is applied tile-by-tile and the outputs are recombined to produce slide thumbnails. Inference times were $22$s for Appendix C Figure 3b and $14$s for Appendix C Figure 3e. We observe the overall consistency of our model outputs with the IHC ground truth. However, false positives, indicated by misplaced DAB signal, remain a problem for our model, and attenuating these will be the subject of future work. Surprisingly, though these test slides belong to the GR dataset, we found the model trained on CAMELYON16 data produced outputs (pictured) at least comparable to those from the GR model, indicating a readily transferable method.  

\section{Discussion}

In this paper we have demonstrated a systematic localisation problem with unsupervised CycleGANs for stain transfer in histopathology tiles, and proposed an improved method using a region-based discriminator, leveraging a library of cell interest regions. We have further shown how datasets of different clinical origin may be successfully combined for learning stain transfer models. The proposed pipeline is a semi-automatic means for extracting additional supervision ``for free'', greatly improving the unsupervised baseline. Although for H\&E data, we still rely on an expert annotation, our experiments revealed exciting possibilities for further automation. Firstly, due to a parsimonious design (only assumptions about cell size and clustering are made), the library building pipeline was equally applicable to both datasets and would likely generalise to others. Secondly, we found datasets could be combined and that a model trained on CAMELYON16 H\&E transfers well to the GR dataset. In the latter case, only annotations from CAMELYON16 have been used, implying reusability of a library once computed on CAMELYON16, a free resource.

\subsubsection{Acknowledgments}
This work was partially supported by the ANR  Hagnodice ANR-21-CE45-0007 and ARC SIGNIT201801286. Experiments have been conducted using HPC resources from the \href{http://mesocentre.centralesupelec.fr/}{“Mésocentre”} computing center of CentraleSupélec and École Normale Supérieure Paris-Saclay supported by CNRS and Région Île-de-France. 

% We observe also the potential for RoI discrimination as a data augmentation strategy. One need only program a contour generator to produce contour template images as a source domain, and relate it to some target domain using a CycleGAN. In biomedical image applications in particular, contour generation is often straightforward, as objects of interest (such as cells) adhere to simple geometric shapes (e.g., \cite{hou2019robust}). In effect, the effort of manual annotation could be replaced by the less burdensome effort of designing an image processing pipeline to obtain the library of cell bounding boxes.

% \subsubsection{Acknowledgements} This work was supported by the ARC Grant SIGNIT201801286 and LabEx DIGICOSME scholarship (RD N$^o$ 264). We would like to thank the Mesocenter\footnote{\href{http://mesocentre.centralesupelec.fr/}{http://mesocentre.centralesupelec.fr/}} of CentraleSup\'elec for their computational resources.





% LNCS Homepage, \url{http://www.springer.com/lncs}. Last accessed 4
% \end{thebibliography}

\bibliography{IEEEabrv,2308}

% \begin{figure*}[t!]%
% \centering
% \subfloat[Input]{\includegraphics[width=0.29\textwidth]{img/hes_input_img.png}} \qquad
% \subfloat[Hematoxylin]{\includegraphics[width=0.29\textwidth]{img/hes_ihc_h.png}} \qquad
% \subfloat[Clusters]{\includegraphics[width=0.29\textwidth]{img/hes_clusters.png}} \qquad
% \subfloat[Annotation]{\includegraphics[width=0.29\textwidth]{img/hes_mask.png}} \qquad
% \subfloat[Combined mask]{\includegraphics[width=0.29\textwidth]{img/hes_full_mask.png}} \qquad
% \subfloat[Output]{\includegraphics[width=0.29\textwidth]{img/hes_final.png}}
% \caption{Library-building pipeline for a sample HES tile. FOR SUPPLEMENTARIES}
% \label{fig:hes_library}
% \end{figure*}

% \begin{figure*}[t!]%
% \centering
% \subfloat[Input]{\includegraphics[width=0.29\textwidth]{img/ihc_input_img.png}} \qquad
% \subfloat[Hematoxylin]{\includegraphics[width=0.29\textwidth]{img/ihc_ihc_h.png}} \qquad
% \subfloat[DAB]{\includegraphics[width=0.29\textwidth]{img/ihc_ihc_d.png}} \qquad
% \subfloat[Otsu threshold]{\includegraphics[width=0.29\textwidth]{img/ihc_mask.png}} \qquad
% \subfloat[Closing]{\includegraphics[width=0.29\textwidth]{img/ihc_full_mask.png}} \qquad
% \subfloat[Output]{\includegraphics[width=0.29\textwidth]{img/ihc_final.png}}
% \caption{Library-building pipeline for a sample IHC tile. FOR SUPPLEMENTARIES.}
% \label{fig:ihc_library}
% \end{figure*}

\end{document}